\pdfoutput=1
\documentclass{mem}
\usepackage{natbib}
\usepackage{txfonts}
\usepackage{balance}
\usepackage{graphicx}
\usepackage[a4paper]{hyperref}

\begin{document}
\title{The diffuse X-ray emission from the Galactic center with Simbol-X}
\author{R. \,Belmont\inst{1}  \and M. \, Tagger\inst{2} }
\institute{Centre d'Etude Spatiale des Rayonnements, 9 rue du Colonel Roche, BP44346,  31028 Toulouse Cedex 4, France, \email{belmont@cers.fr} \and
CEA Service d'Astrophysique, UMR "AstroParticules et Cosmologie", Orme des Merisiers, 91191 Gif-sur-Yvette, France
}
\authorrunning{Belmont }
\titlerunning{Diffuse X-ray emission from the GC}
\abstract{
Similarly to the larger Galactic ridge, the Galactic center region presents a hard diffuse emission whose origin has been strongly debated for the past two decades: does this emission result from the contribution of numerous, yet unresolved, discrete point sources ?  Or does it originate in a truly diffuse, hot plasma ?  

The Galactic center region (GC) is however different on many respects from the outer parts of the Galaxy, which makes the diffuse emission issue at the Galactic center unique. Although recent observations seem to favour a point sources origin in the far Galactic ridge, the situation is still unclear at the GC and new observations are required. 

Here we present new results on the modeling of the truly diffuse plasma. Interestingly, such a plasma would strongly affect the dynamics of orbiting molecular clouds and thus the central engine activity. Discriminating between the two hypothesis has thus become a crucial issue in the understanding of this central region that makes the link between the inner small accretion disk and the large scale Galactic dynamics. We investigate  the new inputs we can expect from Simbol-X on this matter.

}
\maketitle{}

\section{Introduction}

For two decades, X-ray observations with {\it Einstein, HEAO, Ginga, ASCA}, and now {\it XMM-Newton} and {\it Chandra} have reported a diffuse emission from the Galactic ridge. This emission extends out to more than 4 kpc in radius ($25^{\rm o}$ with a 8~kpc sun-Galactic center distance) from the central super massive black hole Sgr A*, but in the 1-10 keV band, it is very peaked in the first $2^{\rm o}$ (250 pc) at the Galactic center. The typical spectrum from this central region results from the contribution from several phases \citep{Muno04,Koyama07}. 
\begin{enumerate}
\item[1-] Many K-$\alpha$ and K-$\beta$ lines of very ionized metals (Si, S, Ar, and Ca) are observed in the 1-4 keV band. Those lines have been attributed to the soft ($k_BT\sim 0.8$ keV) plasma of young supernova remnants. 
\item[2-] A fluorescence iron line at 6.4 keV, resulting from the interaction of cosmic rays with cold gas ($k_bT\sim$50 K), is also present. 
\item[3-] The diffuse emission is mainly characterized by two iron lines at 6.7 and 6.9 keV and by the underlying continuum. These lines correspond to very ionized states of Fe and therefore cannot originate in the soft phase. Two main ideas have been suggested to account for these lines. a) They originate in the hot plasma of many unresolved discrete point sources (mostly CVs). b) They originate in a hot ($k_BT\sim 7 $ keV), truly diffuse plasma that bathes the emitting region.
\item[4-] Above 8 keV, the spectrum is line free. The continuum cannot originate in a thermal 7 keV plasma and rather corresponds to a non thermal emission. Again, its not clear whether these processes take place in discrete points source or in a diffuse medium.
\end{enumerate}

Here we address the issue of the diffuse emission at the GC. In the first section, we remind briefly the key points that are common to the Galactic Ridge emission (see the proceeding by S Mereghetti for more details).  In the second section, we describe more precisely the central region and the truly diffuse plasma issue. In the last section, we detail the inputs Simbol-X might bring on these questions.

\section{Origin of the Galactic Ridge and Galactic center diffuse emissions}

Since the first space-based X-ray observations in the 80s, more and more sources have been resolved by the increasing power of observatories. A cosmic contribution from extra-galactic sources was first identified. Galactic sources were also detected and it is now estimated that 85\% of the X-ray emission above 20~keV is resolved in the Galactic ridge \citep{Lebrun04}. The emission at lower energy is however unclear. The typical spectrum is very similar to that of sources such as Cataclysmic Variables, but so far only a fraction of the total X-ray emission in the 1-10~keV band has been resolved. To account for undetected sources, Log(N)-Log(S) diagrams are built and extrapolated to weak sources. This work is subject to possibly large biases and uncertainties. Source confusion can also limit the detection capabilities of the instruments, especially at high energy.

%As the population of weak X-ray sources as CVs were too small to account for the total X-ray emission, the contribution of point sources has long been debated. 
Recent deep observations combined with new estimates of the unresolved population seem to show that the fraction from discrete point sources  could be large in the Galactic ridge \citep{Revnivtsev06}. However, observations with Suzaku have shown that the remaining diffuse emission has not the same profile as the point sources in the first fractions of degrees at the Galactic center \citep{Koyama07}. The point sources may thus contribute differently in the Galactic ridge and at the Galactic center. 

The idea of a diffuse plasma was early suggested since the fits with such a thermal plasma match best the spectrum, but it was also very debated since it raises severe paradoxes. The temperature of this plasma is so high that is should not be confined by the gravitational potential, so that the power required to heat it before it leaves the Galactic plane exceeds any known energy source. Also, even if the plasma was confined, no heating mechanism had been identified. However, as we show in the next section, recent advances in the modeling of a truly hot plasma have solved its paradoxes at the Galactic center. 

%The situation is still unclear and new observations are required. In addition, it has been shown that the existence of a diffuse plasma would have strong implications on the cold gas dynamics in this region. It has thus become crucial  to discriminate between the two origins. 

\section{Possibility of a truly diffuse plasma at the Galactic center}

As can be observed at all wavelengths, the GC region is a very particular region. 1- The X-ray emission is much stronger than in the rest of the Galactic ridge. 2- Infrared observations show that the GC corresponds to a strong concentration of cold molecular gas. At a radius of about 150~pc, the mean density jumps by a factor of 20 from outside to inside. All this cold molecular content is condensed in clouds and forms the so-called {\it Central Molecular Zone} \citep{Morris96,Oka01} that well matches the central X-ray emitting region. 3- At radio wavelengths, observations also show numerous non thermal filaments that are observed nowhere else in the Galaxy \citep{Larosa00}. These filaments are beautifully aligned with the direction perpendicular to the galactic plane and suggest a strong, vertical magnetic field \citep{Morris96}. All these unique features probably result from the same global galactic dynamics and may contribute to make the situation at the GC very different from that farther out in the Galaxy.
And the problems of a diffuse plasma at the GC, (confinement and heating) can be solved by a consistent interaction of these different phases. 

\subsection{Plasma confinement}
The idea that a diffuse 7 keV plasma should escape from the Galactic plane directly results from the (mono-) fluid assumption: the global sound speed of the plasma is larger than the escape velocity required for bodies to leave the gravitational wheel. 
%This indeed could imply that the whole plasma must escape if it where composed by only one species or if all constituting species had the same behavior.

However, the plasma is composed by many species (electrons, protons, He ions and other heavy ions) and, on the typical escape time ($\tau_{\rm esc}\sim 5\times10^4$~yr), the plasma is not collisional ($\tau_{\rm coll} \approx 10^{5}$~yr). This basically means that the different species can have different behaviors and must be studied separately. Such an analysis is very similar to that in planetary atmospheres (except that the gas is a ionized plasma) and the results are comparable. Namely, by comparing the effective thermal velocity of an ion and its electrons ($v_{\rm th} = \sqrt{\frac{k_BT}{\mu m_p}}$, where $\mu$ is a mean molecular weight) with the escape velocity ($v_{\rm esc}\approx$1200~km/s), it is found that:
\begin{itemize}
\item The thermal velocity of protons ($v_{\rm th} \approx$1300 km/s) is larger than the escape velocity. At this hot temperature, protons are too light to be confined by gravity. This is basically the same result as the fluid one.
\item The thermal velocities of heavier ions ($v_{\rm th} <$~750 km/s) are smaller than the escape velocity. Even with this high temperature, any ion other than a proton is heavy enough to be confined by gravity
\end{itemize}
As a consequence, the Galactic center can undergo a selective evaporation that naturally leads to the formation of a heavy, He-dominated plasma, that is confined by the Galactic potential in this region \citep{Belmont05}. 

\subsection{Plasma heating}
If there is no other cooling mechanism, such a bound plasma only cools by radiation, which is much more reasonable since the total X-ray luminosity of the central region is only $L\approx 4\times 10^{37} erg/s$. Such a power can actually be balanced by viscous heating. The numerous molecular clouds that flow in the central region represent a large reservoir of kinetic and gravitational energy that can be dissipated by the viscosity of the diffuse plasma. 

Because of its high temperature, the diffuse plasma is highly viscous ($\nu \propto T^{5/2}$, \citep{Spitzer62}). However, the strong magnetic field inhibits the usual shear viscosity, only leaving the other component: the so called bulk viscosity related to the compression of the fluid. As the clouds motion is rather slow (subsonic), the associated compression is weak, which limits the dissipation efficiency. The overall dissipation depends on the exact wake of the clouds. 

The wake of conducting bodies in a magnetized plasma has been studied in space science to investigate the motion of satellites in space magnetospheres. In first approximation, it is dominated by Alfv\'en perturbations that propagate along the field lines. These Alfv\'en waves forms a wing structure and carry a strong energy flux away from the cloud \citep{Drell65,Neubauer80}. This flux depends on the magnetic field strength and for any value in the expected range (10$\mu$G-1mG), the cumulative flux associated to the motion of all the clouds in the central region is much larger that the X-ray luminosity. Only a fraction of this power is required to heat the diffuse plasma. To first order, Alfv\'en wave cannot be dissipated by the bulk viscosity since they are not compressible. However, it has been shown that non linear effects or a significant curvature of the field line ($R_c \sim$ 100 pc) provide sufficient dissipation to account for the hot plasma temperature \citep{Belmont06}. 

\section{Observations with Simbol-X}
It has long been though that such diffuse plasma could simply not exist in the Galactic plane, so that the only observational challenge was to make deeper observations and improve the knowledge on the weak sources population in order to find enough of them. Observational inputs of Simbol-X for the detection of weak point sources are discussed in a different talk (see S. Mereghetti's proceeding). We emphasize that the identification capabilities of Simbol-X at high energy ($k_BT > 10$~keV) will be even more crucial at the Galactic center at where the source confusion is high. 
On the other hand, the recent results presented here show that a diffuse plasma at $k_BT \sim 7$~keV can survive at the Galactic center and that its hot temperature can be explained by an efficient viscous dissipation. Thus, they give new observational diagnostics that must be investigated as well. 

At 7 keV, hydrogen and helium are fully ionized, so that there is no direct check for the dominant species. Re-interpretation of the X-ray data show that assuming a helium dominated plasma leads to densities and abundances about three times lower than when assuming a hydrogen dominated plasma. The most recent observations with Suzaku give iron abundances about 3.5 solar ones assuming a H-dominated plasma, which gives back solar abundances when assuming a He-dominated plasma.

Also, if there is no strong mixing mechanism, such a gravitationnaly confined plasma should be stratified, the heavier elements concentrating at low latitude whereas light elements extend higher above the Galactic plane. If observed, such a stratification would be a strong evidence for a diffuse hot plasma. 
So far, only the iron line has been observed precisely. The comparison of the scale height of its emission with that of the emission in the He continuum could reveal the stratification. However the spectral region where the iron lines and the continuum are observed is very confused. Contributions from the iron lines (i.e. from the hot phase), from the line complex (i.e. from the soft phase), from the fluorescence line (i.e. from the cold phase) and from the hard possibly non-thermal tail that extends above 10~keV, as well as their respective continuum all add in the same narrow spectral range. As a result, it is difficult to measure with a good accuracy the relative strength of these different components, in particular the He continuum. 

Simbol-X will provide the best spectra of the Galactic center region above 10 keV. This is of first interest for three reasons:
\begin{enumerate}
\item The spectrum above 8 keV is less confused than at lower energy. The direct estimate of the He continuum will thus be easier. 
\item The continuum however remains confused by the flat power law tail extending to higher energy: the thermal continuum of a 7 keV plasma is supposed to drop quickly above 10~keV but observations show that the emission evolves from a thermal to a non-thermal one around 10~keV. Previous high resolution X-ray observatories were not able to get precise spectra above 8-10~keV from the central region and could not get a reliable estimate of the non thermal contribution. Such observations are however critical to measure the cut-off of the thermal emission and it strength in order to characterize the stratification.
\item Given its low background level, Simbol-X will probably be able to resolve the He-like nickel line at 7.8~keV, first observed last year by Suzaku \citep{Koyama07}. This new line corresponds to the emission of a third species constituting the hot plasma. Its detection, compared to the He continuum and the iron line emission may provide more constrains on the stratification.
\end{enumerate}
Several pointings at different latitudes will thus allow for the first time to investigate the stratification of a plasma confined by gravity. 

\bibliographystyle{aa}

\end{document}